\documentclass[review]{elsarticle}

\usepackage{lineno,hyperref}
\modulolinenumbers[5]

\journal{International Journal of Multiphase Flow}

\usepackage{color}
\usepackage{multirow}
\usepackage{dcolumn}
\usepackage{multicol}
\usepackage{graphicx}
\usepackage{caption}
\usepackage{amssymb}
\usepackage{amsmath}
\usepackage{epstopdf}
\usepackage{mathtools}
\usepackage{float}
\usepackage{bm}
\usepackage{ulem}

\usepackage{subcaption}
\usepackage{graphicx}
\usepackage{soul}

\usepackage{color}

\biboptions{authoryear}

\biboptions{sort&compress}

\usepackage{color}
\usepackage{xcolor} 

\def\We{{\it We}}

\def\We{{\it We}}

\newcommand{\ks}{\textcolor{black}} 

\begin{document}

\begin{frontmatter}
\title{Droplet breakup morphologies and the resultant size distribution in an opposed-flow airstream at different Weber numbers}
\author{Shubham Chakraborty$^a$, Someshwar Sanjay Ade$^a$, Lakshmana Dora Chandrala$^b$ and Kirti Chandra Sahu$^c$\footnote{ksahu@che.iith.ac.in}}
\address{$^a$Center for Interdisciplinary Program, Indian Institute of Technology Hyderabad, Kandi - 502 284, Sangareddy, Telangana, India \\
$^b$Department of Mechanical and Aerospace Engineering, Indian Institute of Technology Hyderabad, Kandi - 502 284, Sangareddy, Telangana, India  \\ 
$^c$Department of Chemical Engineering, Indian Institute of Technology Hyderabad, Kandi - 502 284, Sangareddy, Telangana, India}

\begin{abstract}
The present study investigates the morphology and breakup dynamics of a freely falling drop in a vertical airstream using shadowgraphy and in-line holography. The in-line holography provides the temporal evolution of the volumetric size distribution of child droplets formed during various fragmentation processes at different Weber numbers $(\We)$. The droplet undergoes different fragmentation processes at significantly lower Weber numbers in opposed-flow configurations compared to cross-flow configurations. Our findings reveal distinct fragmentation modes, namely bag, bag-stamen, and dual-bag breakup, observed at $\We=9.38$, 16.9, and 18.9, respectively. At $\We = 9.38$, the combined effects of bag rupture, rim breakup, and node fragmentation generate child droplets of varying sizes, driven by the interplay of the Rayleigh-Plateau and Rayleigh-Taylor instabilities. At $\We = 16.9$, the interaction of aerodynamic and shear forces leads to bag-stamen fragmentation, characterized by forming a stamen-like structure along with the bag. Both bag and bag-stamen breakups result in tri-modal size distributions. However, at $\We = 16.9$, fewer tiny droplets are produced compared to the bag breakup observed at lower Weber numbers. In contrast, at $\We = 18.9$, a dual-bag breakup occurs, where both bags inflate and burst simultaneously. This process generates tiny child droplets in the early stages, while larger child droplets form later due to the fragmentation of the rim and nodes, resulting in a bi-modal size distribution. We have performed a theoretical analysis using a two-parameter gamma distribution, which satisfactorily predicts the size distributions observed experimentally at different Weber numbers.
\end{abstract}

\end{frontmatter}


\section{Introduction} \label{sec:intro}

Droplet interactions with an airstream typically occur in cross-flow, opposed-flow and co-flow configurations, which are prevalent in several applications, ranging from industrial processes to natural phenomena \citep{villermaux2007fragmentation,jain2019secondary,villermaux2020fragmentation,hopfes2021secondary,raut2021microphysical,xu2022droplet,kant2022bags,traverso2023data,balla2020numerical,balla2019shape,agrawal2017nonspherical,agrawal2020experimental}. In atmospheric flows, such interactions significantly influence the morphology and size distribution of raindrops, which are critical for accurate rainfall prediction \citep{marshall1948distribution,patade2015particle,chakraborty2025drop}.

A droplet falling vertically under gravity and encountering a horizontal airstream (cross-flow configuration) is subjected to shear forces that cause deformation of the droplet and may lead to its breakup. In this configuration, the droplet traverses the outer layer, shear layer, and potential core of the flow field, spending only a brief period in the potential core due to its orthogonal motion with the airstream. As the droplet accelerates with the airstream, the relative velocity between the two gradually decreases over time. In contrast, in opposed-flow and co-flow configurations, the droplet interacts with an airstream that aligns with its motion. In an opposed-flow configuration, the airstream moves in the opposite direction to the motion of the droplet, while in a co-flow system, both the airstream and the droplet move in the same direction \citep{inamura2009visualization,Villermaux2009single}. In the oppose-flow configuration, the droplet remains within the potential core and moves opposite to the airflow, resulting in a constant aerodynamic drag force acting on the droplet. The dynamics of a droplet in an airstream are governed by the interplay between aerodynamic and surface tension forces, characterized by the Weber number, $\We \equiv {\rho_a U^2 d_0 / \sigma}$, where $\rho_a$ denotes the air density, $\sigma$ is the interfacial tension, $U$ is the \ks{average} velocity of the airstream, and $d_0$ represents the diameter of the droplet. Notably, the critical Weber number ($\We_{cr}$) required for bag fragmentation in the opposed-flow configuration is significantly lower than that in the cross-flow configuration.

In cross-stream configurations, it is well established that a droplet undergoes vibrational breakup at low Weber numbers, characterized by shape oscillations at a specific frequency \citep{taylor1963shape}. As the oscillation amplitude grows, the droplet eventually fragments. At higher Weber numbers, the droplet transitions to forming a single bag structure on its leeward side, encased by a thicker liquid rim. The subsequent rupture of the bag and fragmentation of the rim result in the generation of smaller child droplets, a process commonly referred to as bag breakup \citep{taylor1963shape, kulkarni2014bag}. This bag fragmentation process is driven by the Rayleigh-Taylor (RT) instability, the Rayleigh-Plateau (RP) capillary instability \citep{taylor1963shape}, and the nonlinear instability of liquid ligaments \citep{jackiw2021aerodynamic, jackiw2022prediction}. The critical Weber number ($\We_{cr}$) for transitioning from vibrational to bag breakup in a cross-stream configuration is approximately 12 \citep{jain2019secondary,zhao2010morphological,soni2020deformation,kirar2022experimental}. The bag-stamen and multi-bag breakup modes, which share similarities with the bag breakup mode, involve additional features like the formation of a central stamen, resulting in either a larger ligament (bag-stamen) or several smaller droplets (multi-bag) during fragmentation. For intermediate Weber numbers ($28 \le \We \le 41$), the fragmentation leads to the formation of multiple bags \citep{cao2007new}. At higher Weber numbers, the droplet exhibits a shear mode, where the edge of the droplet deflects downstream, causing the liquid sheet to fracture into tiny droplets. At very high Weber numbers, the droplet rapidly disintegrates into a cluster of small fragments, resulting in catastrophic fragmentation. All these droplet breakup phenomena in a cross-flow configuration have been investigated by several researchers over the last several decades \citep{pilch1987use,guildenbecher2009secondary,fakhari2011investigation,flock2012experimental,gao2013quantitative,kekesi2014drop,kulkarni2014bag,xiao2017simulation,yang2017transitions,jain2019secondary,kulkarni2023interdependence,niranjan2024towards}. Recent studies by \citet{ade2023size, joshi2022droplet, boggavarapu2021secondary} have demonstrated various breakup modes, including bag, bag-stamen, dual-bag and multi-bag, across a range of Weber numbers in the cross-stream configuration. \citet{ade2024prf} also investigated the effect of droplet dispensing height on shape oscillations, revealing the intricate interplay between inertia and surface tension forces, which significantly influences radial deformation and breakup dynamics under identical airstream conditions (for a fixed Weber number). \citet{kirar2022experimental} examined the fragmentation of a freely falling droplet in a horizontal swirling airstream at a fixed Weber number using shadowgraphy, identifying a new breakup mechanism called the retracting bag breakup mode. In this mode, the swirling airstream stretches the ligaments in opposite directions, inducing capillary instability and causing the droplet to fragment. \citet{ade2023size,ade2022droplet} employed the in-line holography technique to investigate the size distribution of the child droplets associated with these breakup modes.

In a counter-current airstream, \citet{Villermaux2009single} demonstrated that the droplet undergoes bag fragmentation at a Weber number of approximately 6, which is considerably lower than that observed in the cross-stream configuration. This disparity arises because, unlike in the cross-stream scenario, the droplet in the opposed-flow configuration remains within the potential core region of the airstream throughout the fragmentation process. \citet{inamura2009visualization} employed shadowgraphy to study droplet fragmentation in the opposed-flow configuration, revealing that bag-type breakup is characterized by alternating vortices and bulging, while umbrella-type breakup involves symmetrical vortices and peripheral bulging. It was also observed that the breakup mode transitions from bag to umbrella with a slight increase in relative velocity. \citet{soni2020deformation} investigated the dynamics and breakup of a droplet freely falling under gravity within an oblique airstream oriented at various angles. Their findings revealed that the critical Weber number for bag breakup decreases as the airstream orientation shifts from cross-flow to opposed-flow, approaching a value of 6 when the airstream angle with the horizontal exceeds $60^\circ$. They also observed that, in an oblique configuration, the droplet follows a curvilinear motion while undergoing topological changes. Additionally, the critical Weber number was found to be influenced by factors such as the initial droplet size, the fluid properties of the liquid, the ejection height from the nozzle and the velocity profile of the airstream.

In early studies, various experimental techniques, such as planar methods, imaging techniques, laser-induced fluorescence, and femtosecond pulsed light sources, were employed to investigate droplet breakup \citep{tropea2011optical,rajamanickam2017dynamics}. \citet{boggavarapu2021secondary} analyzed droplet size distributions associated with different breakup modes using the particle/droplet image analysis (PDIA) method. They found that bag and bag-stamen breakups resulted in a tri-modal size distribution, while dual-bag and multi-bag breakups produced a bi-modal distribution. Recently, digital in-line holography has become a powerful tool for estimating droplet size distributions \citep{guildenbecher2017characterization, shao2020machine, radhakrishna2021experimental, essaidi2021aerodynamic, li2022secondary, ade2022droplet, ade2023size}. \citet{radhakrishna2021experimental} examined the effect of the Weber number on droplet fragmentation at high Ohnesorge numbers, exploring various breakup modes using this technique. \citet{ade2023size, ade2022droplet,ade2024prf} applied digital in-line holography to study droplet fragmentation in both straight and swirling horizontal airstreams. In \citet{ade2023size}, the droplet size distribution for different Weber numbers undergoing single-bag and multi-bag fragmentation in a cross-flow configuration was analyzed. They showed that, despite six distinct breakup mechanisms, dual-bag breakup exhibited a bi-modal distribution, whereas single-bag breakup followed a tri-modal distribution. Furthermore, they demonstrated that the analytical model proposed by \citet{jackiw2022prediction} accurately predicts droplet size distributions across a broad range of Weber numbers.

As the above review indicates, while the fragmentation of droplets in a horizontal airstream has been extensively studied for several decades, droplet interactions with a vertical airstream in the opposed-flow configuration have received far less attention \citep{inamura2009visualization,Villermaux2009single}, despite their significance in accurate weather predictions, combustion, surface coating, pharmaceutical production, disease transmission modelling, artificial rain technology, and numerous other applications \citep{ellis1997effect,lefebvre2017atomization}. In the present study, we investigate the dynamics of droplet breakup in an opposed-flow airstream for different Weber numbers by conducting experiments using shadowgraphy and in-line holography techniques, enabling a more comprehensive characterization of the size distribution of the child droplets resulting from fragmentation. In contrast, previous research has often relied only on shadowgraphy, which does not offer detailed insight into the size distribution of child droplets resulting from the fragmentation. Additionally, we conduct Particle Image Velocimetry (PIV) experiments to gain insights into the flow field of the airstream and perform a theoretical analysis using the model proposed by \citet{jackiw2022prediction}, which accurately predicts the size distributions observed in our experiments. 

The rest of the paper is organized as follows. The experimental setup and procedure are described in \S\ref{sec:expt}. The experimental results on droplet fragmentation in a counter-current airstream and the resulting size distribution are presented in \S\ref{sec:dis}. This section also introduces the theoretical model to predict the size distributions of child droplets from various breakup modes at different Weber numbers. Further, the results from the theoretical model are compared with the experimental findings. Finally, the conclusions are presented in \S\ref{sec:conc}.

\section{Experimental procedure}\label{sec:expt}

In the present study, we investigate the morphology and breakup phenomena of a freely falling droplet in a continuous airstream with an opposed-flow configuration, along with the resulting size distribution of the child droplets, using shadowgraphy and machine learning-based digital in-line holography techniques. The strength of the airstream is varied by adjusting the flow rate from an air nozzle, which in turn alters the Weber number encountered by the droplet, leading to fragmentation. The average velocity of the vertical airstream ($U$) is maintained at 12.45 m/s, 16.7 m/s, and 17.68 m/s, which correspond to Weber numbers ($\We$) of 9.38, 16.9, and 18.9, respectively. Although the droplet is slightly deformed when dispensed from the needle, we assumed it to be spherical for the calculation of Weber numbers, approximating it as an equivalent spherical droplet with an initial diameter of $d_0 = 3.6 \pm 0.07 ~ \text{mm}$. The airstream for different Weber numbers is characterized using the Particle Image Velocimetry (PIV) technique. 

\begin{figure}
\centering
\includegraphics[width=0.85\textwidth]{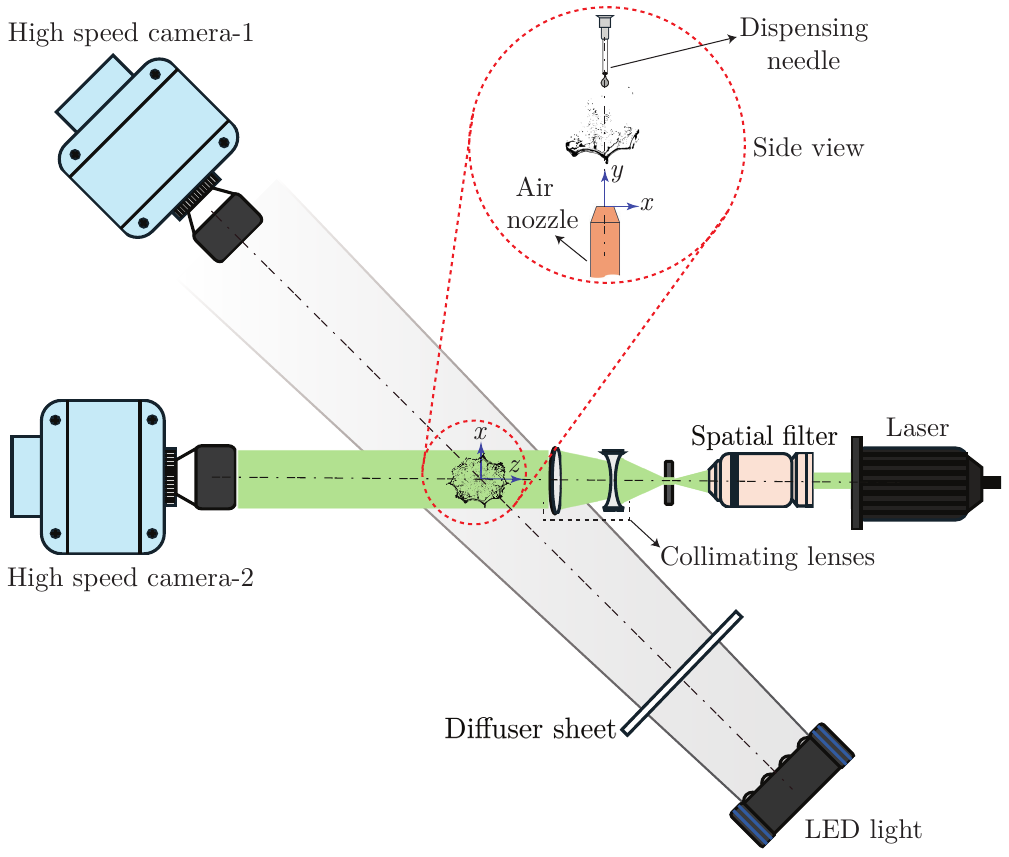}
\caption{
Schematic of the experimental setup (top view) equipped with shadowgraphy and digital inline holography techniques, used to investigate the size distribution of child droplets resulting from the fragmentation of a freely falling water droplet exposed to a vertical airstream in an opposed-flow configuration. The setup includes two high-speed cameras, a spatial filter, collimating lenses, a laser, a diffuser sheet and a light source. The inset illustrates the arrangement of the dispensing needle and air nozzle, as well as the droplet fragmentation occurring between the needle and the air nozzle.}
\label{exp_setup}
\end{figure}

A schematic of the experimental setup, depicting its top view, is presented in figure \ref{exp_setup}. \ks{The setup consists of (i) an air nozzle with an inner diameter of $D_{n} = 18$ mm, (ii) a droplet dispensing needle (18 gauge), (iii) a continuous-wave laser, having an output power 100 mW and wavelength 532 nm (SDL-532-100 T, Shanghai Dream Lasers Technology Co. Ltd), (iv) a spatial filter arrangement consisting of an infinity-corrected plan achromatic objective (20× magnification; Holmarc Opto-Mechatronics Ltd.), (v) collimating optics with concave and convex lenses (Holmarc Opto-Mechatronics Ltd), (vi) two high-speed cameras (Phantom VEO 640L; Vision Research, USA), (vii) a high-power light-emitting diode (MultiLED QT, GSVITEC, Germany), (viii) a syringe pump (model: HO-SPLF-2D; make: Holmarc Opto-Mechatronics Pvt. Ltd, India) to generate the same size of liquid droplets and (ix) a diffuser sheet.}

\ks{The experiments were conducted using air and distilled water as the working fluids at an ambient temperature of $25^\circ$C. The physical properties of water considered in this study are density ($\rho_w = 998$ kg/m$^3$) and viscosity ($\mu_w = 1.0$ mPa$\cdot$ s). The density of air ($\rho_a$) is taken as 1.225 kg/m$^3$. The interfacial tension of the air-water interface is $\sigma = 0.072$ N/m.} These fluid properties are used for modeling and analysis throughout the study. A Cartesian coordinate system $(x, y, z)$ is used to describe the flow dynamics, with the origin located at the center of the air nozzle, as shown in the inset of figure \ref{exp_setup}. The acceleration due to gravity $(g)$ acts in the negative $y$ direction. The dispensing needle is positioned at $(x/D_{n}, y/D_{n}, z/D_{n}) = (0.0, 50.0, 0.0)$. To capture the morphology of the droplet undergoing fragmentation, shadowgraphy is employed using high-speed camera 1, positioned at $x = 500$ mm and oriented at an angle of $45^\circ$ with respect to the $x$ axis. A high-power light-emitting diode, along with a uniform diffuser sheet, provides background illumination. Images are captured at 1600 frames per second (fps) with an exposure time of 1 $\mu$s. In the shadowgraphy, the spatial resolution is 52.59 $\mu$m/pixel, and the image resolution is $1024 \times 600$ pixels.

\begin{figure}
\centering
\includegraphics[width=0.8\textwidth]{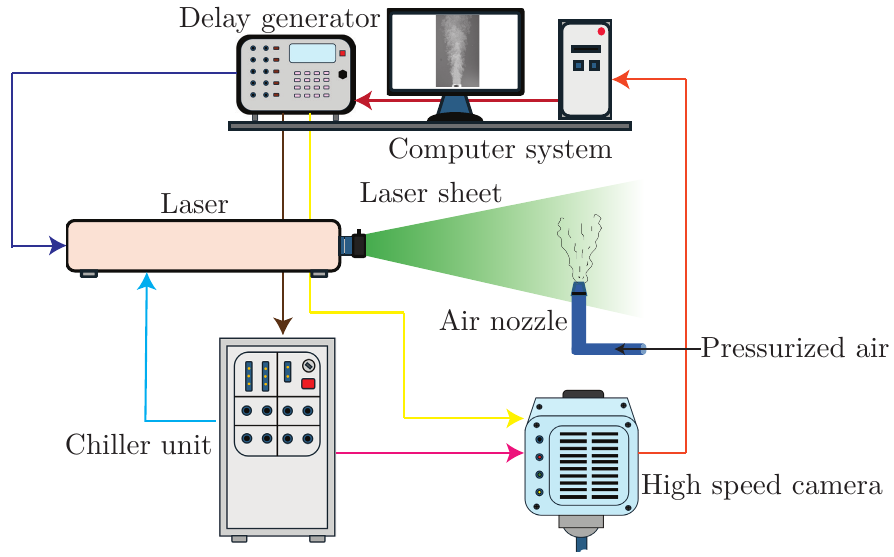}
\caption{Schematic of the Particle Image Velocimetry (PIV) setup used to visualize the flow field and measure the local velocity, which is subsequently employed to calculate the local Weber number for droplet fragmentation.}
\label{PIV_set}
\end{figure}

In digital in-line holography, as shown in figure \ref{exp_setup}, a collimated, coherent, and expanded laser beam is used to record interference patterns. These patterns result from the interaction between scattered light from droplets (object wave) and unscattered background illumination (reference wave) on a camera sensor. The recorded hologram captures both the amplitude and phase information of the object wave. This technique uses a single beam as both the reference and object beam, making it compact and efficient. A continuous-wave laser with an output power of 100 mW and a wavelength of 532 nm is employed. The main components of the setup include a spatial filter, collimating lenses and high-speed camera 2, which is positioned at $x = 500$ mm. The spatial filter consists of an infinity-corrected plan achromatic objective (20X magnification) and a 15 $\mu$m pinhole, ensuring a clean beam. This beam is expanded using a plano-concave lens and collimated by a plano-convex lens to uniformly illuminate the droplet field of view. High-speed camera 2 records the resulting interference patterns at a resolution of $896 \times 800$ pixels, operating at 1800 fps with an exposure time of 1 $\mu$s and a spatial resolution of 48.36 $\mu$m/pixel. Morover, we employ a machine-learning-based post-processing approach to approximate the three-dimensional structure of objects with high spatial resolution, facilitating the determination of the spatial distribution of child droplets and the estimation of their sizes obtained from digital inline holography \citep{ade2024application}. In the present study, approximately 100 manually annotated (ground truth) images from the reconstructed volume and around 500 synthetic holograms are used for network training. The annotations are generated using local thresholding around each child droplet. These manually annotated images, derived from various repetitions, enable the use of a single set of training weights across all experiments. Prior to training, we perform data augmentation by rigidly and elastically deforming the ground-truth images, reducing the need for large datasets. The manual annotations or masks are created by maximizing edge sharpness. A detailed description of the digital in-line holography technique and the associated post-processing method, which incorporates machine learning, can be found in \citet{ade2023size,ade2022droplet}.


The schematic of the Particle Image Velocimetry (PIV) setup used to characterize the airstreams for different Weber numbers is shown in figure \ref{PIV_set}. The PIV technique enables precise measurement of air velocity at various locations within the flow field, which is essential for calculating the Weber number. The setup includes an Nd:YAG laser \ks{(Litron Nano L200-15-LM3128)}, a high-speed camera \ks{(Phantom VEO 640L; Vision Research, USA)}, an air nozzle, a delay generator \ks{(610036, TSI, USA)}, a chiller unit and a computer for data acquisition. The laser generates a thin laser sheet that illuminates the flow field, where small, neutrally buoyant tracer particles with diameters ranging from 1 to 3 $\mu$m are introduced into the airflow using a Laskin sprayer. These particles follow the airstream, providing accurate flow visualization. The particle concentration is adjusted by modifying the boost pressure of the Laskin sprayer, ensuring optimal seeding density for the PIV measurements. The double-pulse laser fires at a 0.05 s interval to illuminate the tracer particles. The high-speed camera records particle positions at a resolution of 2560 $\times$ 1600 pixels and a frame rate of 1000 fps. A delay generator synchronizes the laser pulses with the camera exposure, ensuring precise capture of particle displacement for velocity calculation. The camera is calibrated and positioned perpendicular to the laser sheet to obtain an accurate view of the flow. 

\ks{The raw images obtained from the experiment were processed using PIVlab,  an open-source MATLAB-based software \citep{thielicke2021particle}.  The PIV analysis employed a Fast Fourier Transform (FFT)-based cross-correlation algorithm. To enhance spatial resolution, a multi-pass refinement approach combined with a window deformation method was implemented. The first pass employed an interrogation area of $128\times 128$ pixels with a 50\% overlap (step size: 64 pixels), followed by subsequent passes using $64 \times 64$ pixels (step: 32 pixels) and $32 \times 32$ pixels (step: 16 pixels) to refine the velocity field. The physical size of the final interrogation pass is about 2 mm $\times$ 2 mm. A Gaussian $(2 \times 3) -$point sub-pixel estimator was applied to accurately determine the displacement peak. Furthermore, spurious vectors were identified and removed using a combination of velocity limit filtering, a standard deviation filter (threshold: $8 ~\times$ standard deviation), and a local median filter to ensure data reliability.}


\section{Results and discussion} \label{sec:dis}

\begin{figure}
\centering
\includegraphics[width=0.9\textwidth]{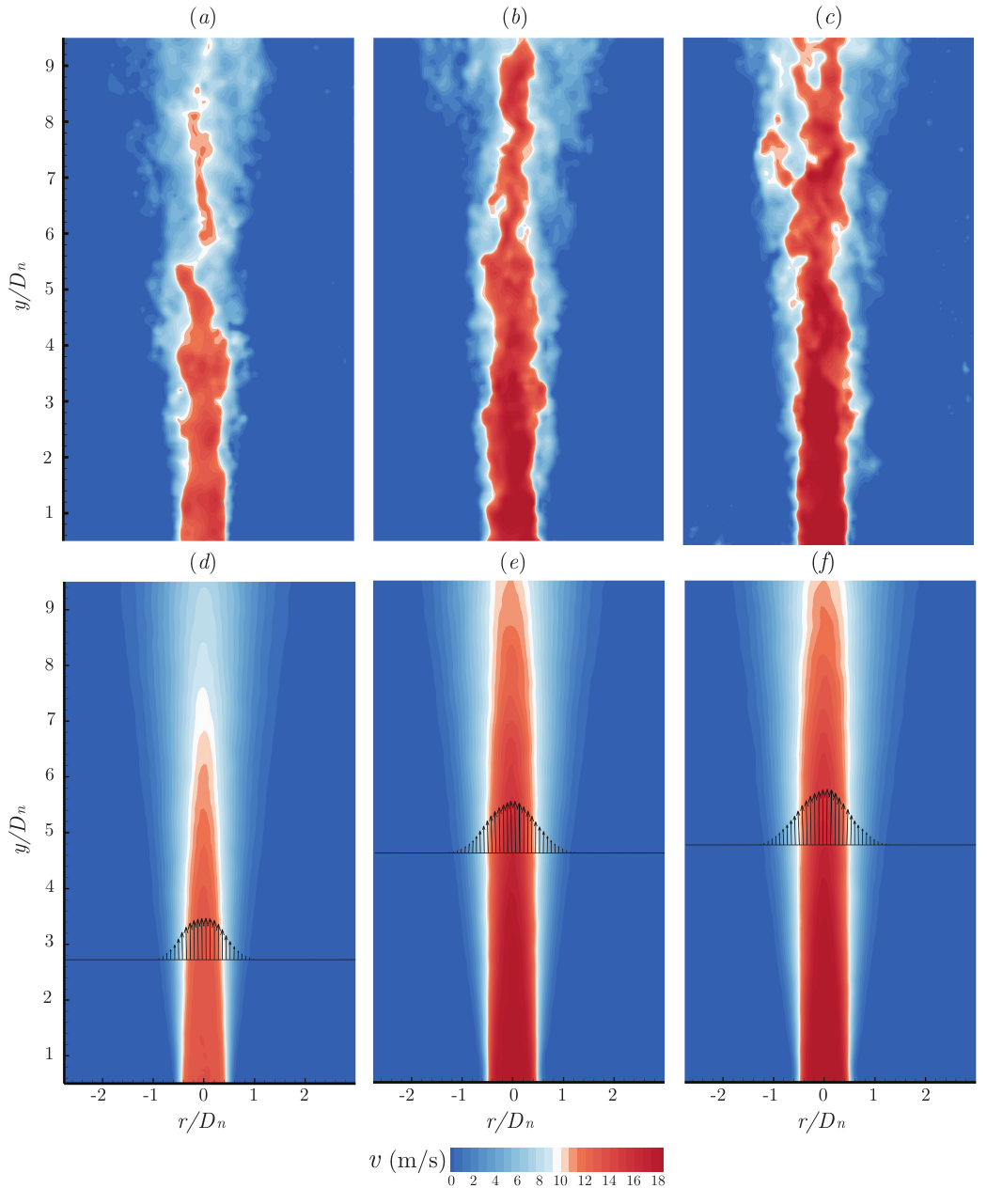}
\caption{Flow visualization of airstreams obtained through the Particle Image Velocimetry (PIV) for different Weber numbers. Panels (a), (b) and (c) depict the instantaneous velocity fields for $\We = 9.38$, 16.9 and 18.9, respectively, while panels (d), (e) and (f) present the corresponding time-averaged velocity fields. The velocity profiles of the airstream at the droplet fragmentation locations, corresponding to different Weber numbers, are overlaid in panels (d), (e), and (f). The color bar represents the velocity magnitude, and $r/D_n = 0$ denotes the axis of the air nozzle. The Weber numbers $\We = 9.38$, 16.9 and 18.9 correspond to airstream velocities of $U = 12.45$ m/s, $U = 16.7$ m/s and $U = 17.68$ m/s, respectively.}
\label{PIV}
\end{figure}

As discussed in the introduction, a droplet falling under gravity and interacting with a horizontal airstream (cross-flow configuration) undergoes distinct fragmentation processes when the Weber number exceeds a critical value $(\We \approx 12)$. In contrast, \citet{Villermaux2009single} demonstrated that a droplet interacting with a vertical airstream experiences bag fragmentation at significantly lower Weber numbers $(\We \approx 6)$. A key difference between these scenarios lies in the interaction duration of the droplet with the potential core region of the airstream. In a cross-flow configuration, the droplet spends less time in the potential core region of the airstream due to its perpendicular trajectory. For instance, \citet{ade2023size} reported that during bag breakup in a cross-flow configuration, the total interaction time of a droplet with the airstream was approximately 26 ms. In contrast, in the present study involving an opposed-flow configuration, we observe that the droplet remains within the potential core region for the entire duration of its flight and breakup ($\sim$ 45 ms). Consequently, this prolonged interaction can significantly influence the fragmentation dynamics of the primary droplet and alter the size distribution of the resulting child droplets at different Weber numbers in the opposed-flow configuration. This aspect, which has not been explored in previous studies, constitutes the primary focus of the present work.

\begin{figure}
\centering
\includegraphics[width=0.65\textwidth]{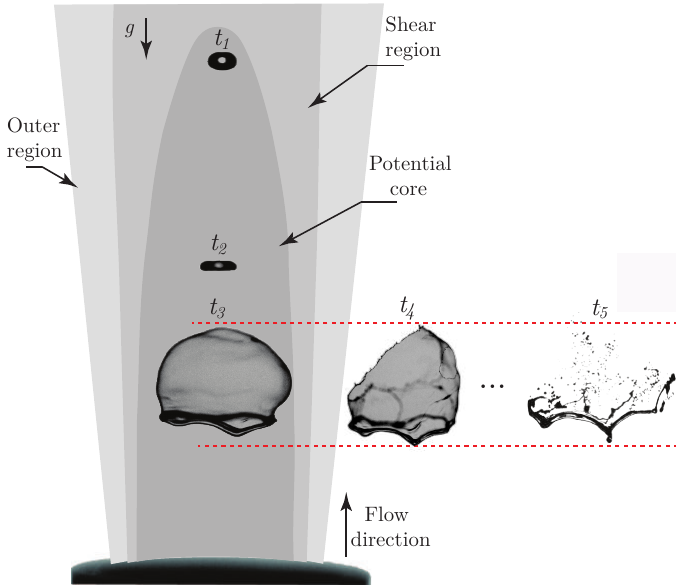}
\caption{Schematic representation of the flow field regions in the airstream in an opposed flow configuration, illustrating the positions of the droplet as it enters the potential core region under the action of gravity at different time instants $(t_1 < t_2 < t_3 < t_4 < t_5)$.}
\label{flow_field}
\end{figure}

The potential core, typically defined as the region where the centerline velocity remains above 95\% of the nozzle exit velocity \citep{martin1977heat, jambunathan1992review}, plays a crucial role in governing droplet breakup characteristics and the resultant distribution of the child droplets. To characterise the airstream, figure \ref{PIV} presents velocity fields obtained using Particle Image Velocimetry (PIV) for different Weber numbers considered in this study. Figure \ref{PIV}(a-c) shows instantaneous velocity distributions for Weber numbers $\We = 9.38$, 16.9 and 18.9, corresponding to airstream velocities $U = 12.45$ m/s, $U = 16.7$ m/s and $U = 17.68$ m/s, respectively. The time-averaged velocity fields, computed from 500 frames captured at 1000 fps, are shown in figure \ref{PIV}(d-f) for the corresponding Weber numbers. These plots also depict velocity profiles at the fragmentation locations for the respective Weber numbers. It can be seen that in the counterflow configuration, due to the entrainment of the surrounding air, the airstream creates a continuously expanding flow field resembling an inverted cone, effectively trapping the descending droplet. We observe that increasing the airstream velocity increases the height of the breakup region from the air nozzle. Specifically, the normalised location of the breakup region, $y/D_n$, is found to be 2.75, 4.63 and 4.82 for $\We = 9.38$, 16.9 and 18.9, respectively. Figure \ref{flow_field} schematically illustrates the three distinct regions within this vertical flow field, namely the outer region, the shear region, and the potential core region. The droplet is dispensed at $y/D_n=50.0$, using a syringe pump aligned with the centerline of the air nozzle. Under the influence of gravity, the droplet enters the potential core region (as shown at $t_1$) and deforms into a disk shape (at $t_2$). The disk-shaped droplet continues to move downward, expanding and forming an inverted bag until it becomes neutrally buoyant. Its thickness reduces as the droplet bulges upward due to the aerodynamic force of the airstream (at $t_3$). Eventually, the droplet undergoes fragmentation at nearly the same vertical location, as shown at $t_4$ and $t_5$ in figure \ref{flow_field}.

In the following discussion, we examine the morphology of a droplet with an initial diameter of $d_{0} = 3.6 \pm 0.07$ mm as it undergoes fragmentation at varying Weber numbers. These Weber numbers are controlled by adjusting the flow rate of the vertical airstream from the air nozzle, as depicted in figure \ref{exp_setup}.

\subsection{Droplet morphology}\label{sec:morp}

\begin{figure}
\centering
\includegraphics[width=0.9\textwidth]{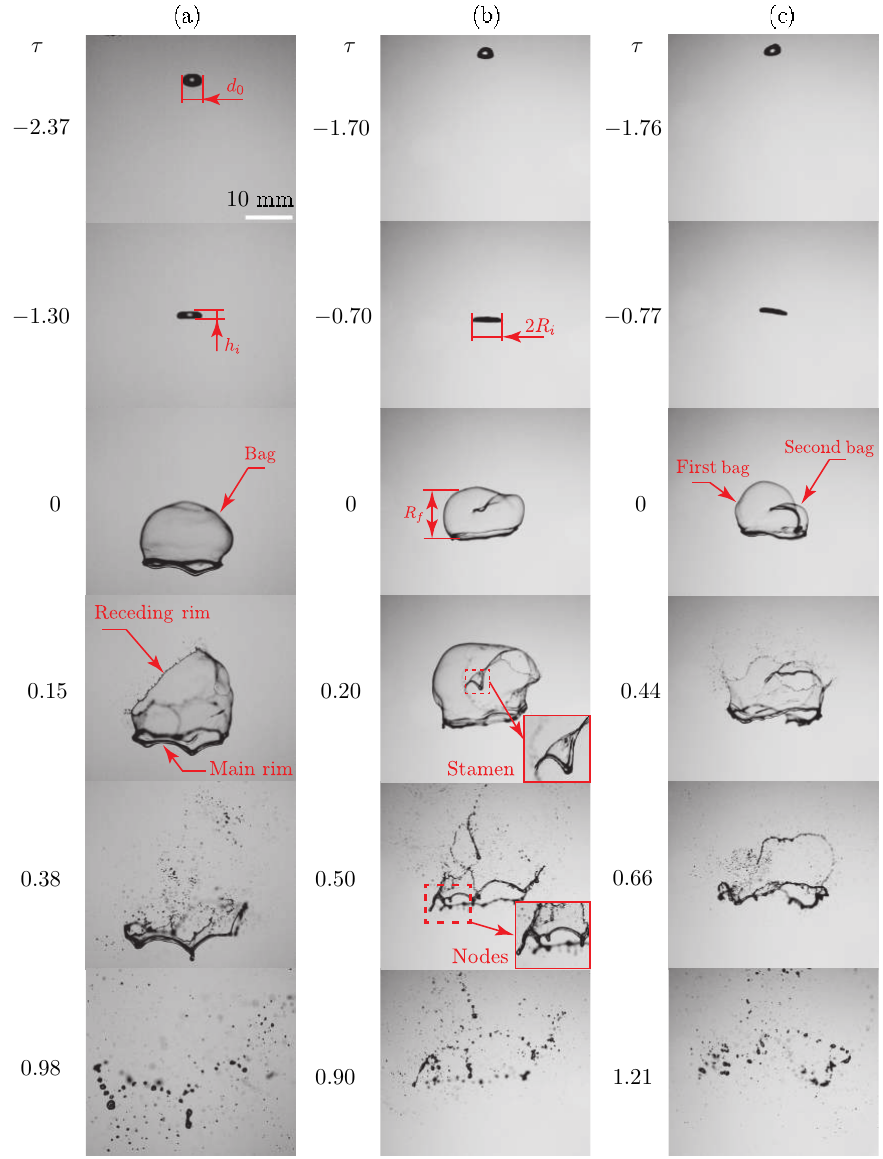}
\caption{Temporal evolution of the droplet breakup dynamics for (a) $\We = 9.38$ (bag breakup), (b) $\We = 16.9$ (bag-stamen breakup) and (c) $\We = 18.9$ (dual bag breakup). The initial diameter of the primary droplet is $d_0 = 3.6 \pm 0.06$ mm. The dimensionless time is defined as $\tau = U t \sqrt{\rho_a/\rho_w} / d_0$, where $U$ is the average velocity of the airstream, $t$ is the physical time, $\rho_a$ is the density of air and $\rho_w$ is the density of water. The instant $\tau = 0$ denotes the instant associated with the onset of droplet fragmentation.}
\label{fig1a}
\end{figure}

Figure \ref{fig1a}(a-c) presents the temporal evolution of droplet fragmentation for $\We = 9.38$, 16.9 and 18.9, respectively. The results are presented in terms of the dimensionless time, $\tau = U t \sqrt{\rho_a / \rho_w} / d_0$, where $\tau = 0$ is defined as the onset of breakup. This dimensionless time is introduced to capture the small characteristic timescale associated with droplet deformation and fragmentation in an airstream \citep{nicholls1969aerodynamic}. Figure \ref{fig1a}(a) illustrates the morphology and breakup dynamics for $\We = 9.38$, characterized by the bag breakup phenomenon. Upon entering the potential core region of the airstream, the droplet deforms into an oblate shape ($\tau = -2.37$) and subsequently flattens into a disk-like form due to aerodynamic forces ($\tau = -1.30$). When the thickness of the droplet reaches a critical value (thickness of the disk-shaped droplet, $h_i$, as shown at $\tau = -1.30$), a thin, inverted bag-like membrane inflates, accompanied by the accumulation of liquid along its periphery, forming a thicker rim. \citet{liu1997analysis} proposed that small holes created by airstream disturbances trigger the rupture of the liquid bag at its tip. However, alternative mechanisms of droplet fragmentation, such as the Rayleigh-Taylor instability, have also been suggested by \citet{Villermaux2009single}. Additionally, \citet{varkevisser2024effect} provided a review on bag breakup mechanisms in sprays, highlighting various hydrodynamic instabilities, including the Rayleigh-Taylor instability. The disintegration of the bag produces tiny child droplets is observed at $\tau = 0.15$. As the bag ruptures, a receding rim forms along the edge of the ruptured membrane, retracting and eventually merging with the primary rim at $\tau = 0.38$. The fragmentation of the rim is driven by the Rayleigh-Plateau instability, which destabilizes the thin liquid structure and leads to the generation of intermediate-size child droplets. Concurrently, the nodes, finger-like structures descending from the rim, fragment due to the Rayleigh-Taylor instability, which arises from the acceleration-induced density stratification between the denser liquid and the lighter air, leading to the formation of larger child droplets. The interplay of the Rayleigh-Plateau and the Rayleigh-Taylor instabilities governs the disintegration of the droplet. Thus, it can be observed that the combined effects of bag rupture, rim breakup, and node fragmentation result in the production of child droplets of different sizes, as evident at $\tau = 0.98$.

Figure \ref{fig1a}(b) shows the morphology and breakup dynamics for $\We = 16.9$, illustrating the bag-stamen fragmentation phenomenon. This type of fragmentation occurs when both aerodynamic and shear effects are significant. When the droplet enters the potential core region of the airstream, it undergoes more deformation, forming a wider disk-shaped drop (see $\tau = -0.70$), compared to the deformation observed at $\We = 9.38$. As time progresses, the droplet deforms further, forming a thin membrane with a thicker rim, similar to the bag breakup mode. However, the increased aerodynamic forces cause non-uniform deformation, leaving an undeformed core at $\tau = 0$. At $\tau = 0.20$, the undeformed core is stretched into a stamen-like protrusion due to aerodynamic drag and shear forces, causing it to elongate and extend outward. While surface tension attempts to restore the spherical shape of the droplet, the aerodynamic forces continue to stretch the stamen, making it thinner and longer until it detaches (see, at $\tau = 0.50$). Simultaneously, the outer rim of the droplet undergoes Rayleigh-Plateau instability, leading to the fragmentation of the rim structure at $\tau = 0.90$. The combined breakup of the stamen and rim results in distinct fragmentation characteristics and size distribution of the child droplets, distinguishing it from the classical bag breakup process (shown for $\We = 9.38$).

Figure \ref{fig1a}(c) demonstrates the morphology and breakup dynamics for $\We = 18.9$ exhibiting the dual-bag fragmentation process. In this scenario, the droplet experiences a much stronger vertical airstream, leading to extensive deformation and the formation of a slightly tilted disk-shaped drop. At $\tau = -0.77$, the droplet deforms into a convex shape due to aerodynamic drag, with a slight tilt that actually represents the undeformed core of the droplet, subsequently leading to a second bag. At $\tau = 0$, the droplet forms the first thin bag with a thicker rim, marking the first stage of bag formation. Simultaneously, the undeformed core of the droplet remains intact and large enough to permit the formation of a second bag, which inflates adjacent to the first. As aerodynamic forces continue to act on the droplet, both bags expand. Between $\tau = 0.44$ and $\tau = 0.66$, both the first and second bags rupture, with the thin sheet of the first bag fragmenting into smaller droplets due to Rayleigh-Plateau instability. Simultaneously, the second bag surrounding the core droplet also undergoes fragmentation, driven by the continuous aerodynamic forces. Finally, at $\tau = 1.22$, the rims and nodes associated with both the first and second bags break up. The simultaneous formation and rupture of these two bags lead to unique fragmentation characteristics, distinguishing the dual-bag breakup process from the bag and bag-stamen breakup modes. Comparing the dual-bag breakup mode in the cross-flow configuration studied by \citet{ade2023size} with the opposed-flow configuration observed in the present study reveals an interesting difference. In the cross-flow case (at $\We = 34.8$), the portion of the droplet exposed to strong aerodynamic forces first inflates into a bag and bursts, followed by the remaining liquid deforming into another bag, which then bursts. In contrast, in the opposed-flow configuration (at $\We = 18.9$, significantly lower than the cross-flow case), both bags inflate and burst simultaneously while the droplet remains within the flow field of the airstream.

In the following section, we analyze the droplet size distribution resulting from the distinct fragmentation processes observed at different Weber numbers.

\subsection{Drop size distribution} \label{sec:DSD}

\begin{figure}
\centering
\includegraphics[width=0.8\textwidth]{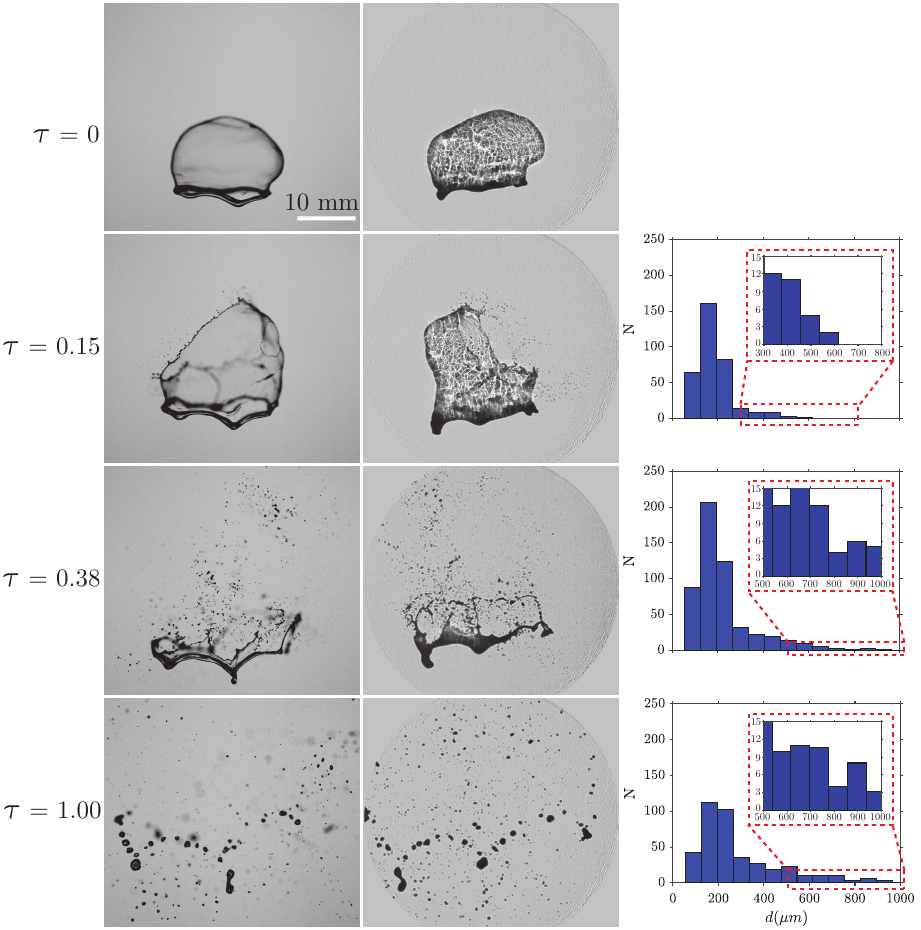}
\caption{Temporal evolution of the droplet size distribution for $\We = 9.38$ (bag breakup phenomenon). The panels in the first and second columns represent the shadowgraphy images and recorded holograms (obtained from the in-line holography technique). The values of the dimensionless time, $\tau$, are mentioned on the left side of the first column, measured from the instant at the onset of the breakup. The scale bar is shown in the top panel of the first column. The panels in the third column depict the histograms of the droplet size distribution (the droplet counts, $N$, versus the droplet diameter, $d$, in $\mu$m) at different instants for $\tau>0$.}
\label{bag}
\end{figure}

Figure \ref{bag} presents the temporal evolution of the droplet size distribution (DSD) for the bag breakup mode at $\We = 9.38$. The first column depicts shadowgraphy images, the second column shows holograms, and the third column displays the count-based histograms of size distributions of the child droplets resulting from the fragmentation. As $\tau = 0$ denotes the onset of bag rupture, the histograms of the droplet size distribution (the droplet counts $(N)$ versus the droplet diameter $(d)$ in $\mu$m) are plotted for $\tau>0$. At $\tau = 0.15$, the bag ruptures due to the combined effects of Rayleigh-Taylor and Rayleigh-Plateau instabilities, driven by aerodynamic forces \citep{taylor1963shape}. This rupture generates tiny child droplets of size ranging from 50 $\mu$m to 700 $\mu$m, with a higher concentration in the 50 $\mu$m to 400 $\mu$m range, resulting in a single peak in the histogram. At $\tau = 0.38$, the ruptured thin sheet of the bag rolls back towards the rim, detaching the attached nodes. This process increases the count of smaller droplets ($50 < d < 300~\mu$m), forming a sharper peak. Additionally, the detachment of larger nodes creates smaller peaks for the $d >500~\mu$m range. This can be clearly observed in the enlarged view of the histogram inset at $\tau = 0.38$. Finally, at $\tau = 1.0$, the rim and nodes disintegrate, producing larger droplets that appear as distinct peaks in the zoomed view of the histogram. Subsequently, no further fragmentation of the child droplets is observed, signifying the completion of the fragmentation process.


\begin{figure}
\centering
\includegraphics[width=0.9\textwidth]{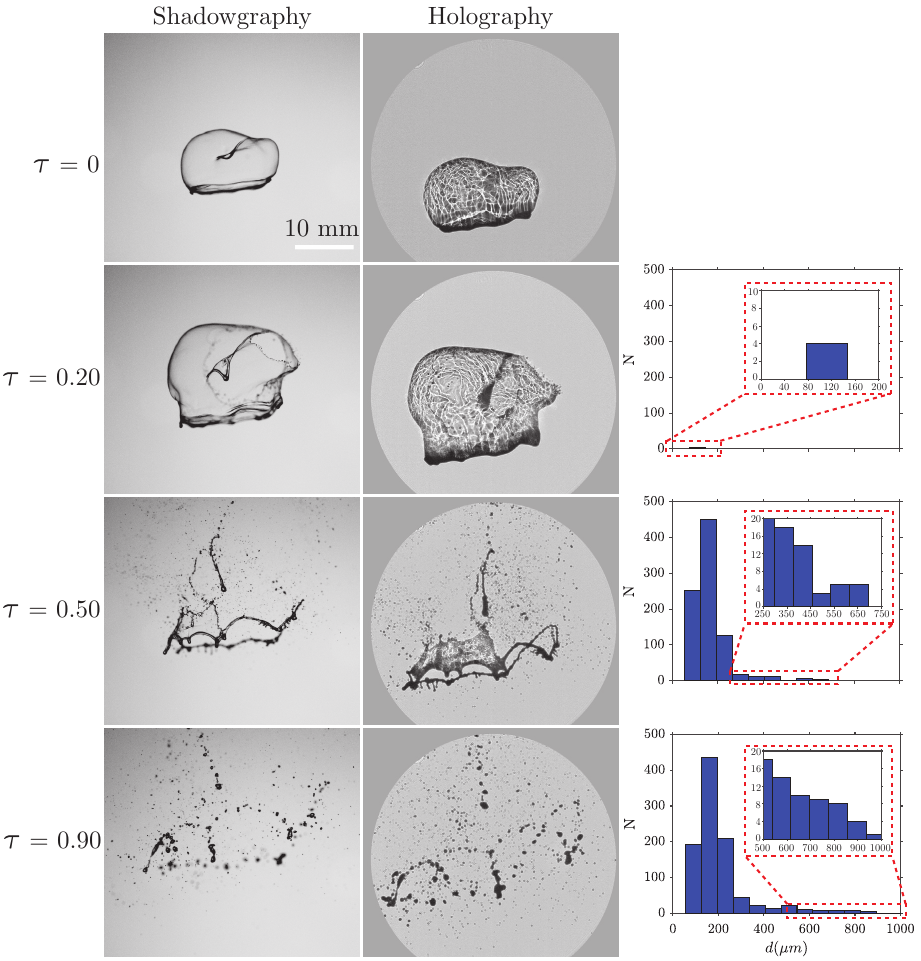}
\caption{Temporal evolution of the droplet size distribution for $\We = 16.9$ (bag-stamen breakup). The panels in the first and second columns represent the shadowgraphy images and recorded holograms (obtained from the in-line holography technique). The values of the dimensionless time, $\tau$, are mentioned on the left side of the first column, measured from the instant at the onset of the breakup. The scale bar is shown in the top panel of the first column. The panels in the third column depict the histograms of the droplet size distribution (the droplet counts, $N$, versus the droplet diameter, $d$, in $\mu$m) at different instants for $\tau>0$.}
\label{bs}
\end{figure}

Figure \ref{bs} presents the temporal evolution of the DSD resulting from the bag-stamen breakup mode at $\We = 16.9$. In this case, it can be observed that the length of the inflated bag is smaller than that of the normal bag fragmentation scenario depicted in figure \ref{bag}. Just after the rupture of the bag ($\tau = 0.20$), it can be seen that, although the bag has ruptured, it produces an almost negligible number of tiny child droplets with $d < 100 ~ \mu$m, a behavior distinct from the normal bag breakup phenomenon. Due to the non-uniform pressure distribution, a relatively small undeformed core remains intact. As the bag is drawn from the deformed sheet by aerodynamic forces, the undeformed portion gets pinched off at its center and stretches to form a stamen \citep{jackiw2021aerodynamic}. As the air continues to accelerate through the inflated bag, it undergoes further fragmentation, producing child droplets ranging from 50 $\mu$m to 750 $\mu$m, corresponding to the prominent peak observed in figure \ref{bs} at $\tau = 0.50$. In the enlarged view, small peaks representing larger droplets ($d > 450~\mu$m) can be observed, which are caused by the detachment of nodes from the bag. Subsequently, the stamen, rim and nodes disintegrate due to capillary instability, resulting in a broader droplet size distribution ranging from 50 $\mu$m to 1000 $\mu$m (see the inset at $\tau = 0.90$). Additionally, larger droplets, though fewer in number, are observed in the enlarged view of the histogram at $\tau = 0.90$ in figure \ref{bs}, originating from the disintegration of the rim, nodes, and stamen.

\begin{figure}
\centering
\includegraphics[width=0.9\textwidth]{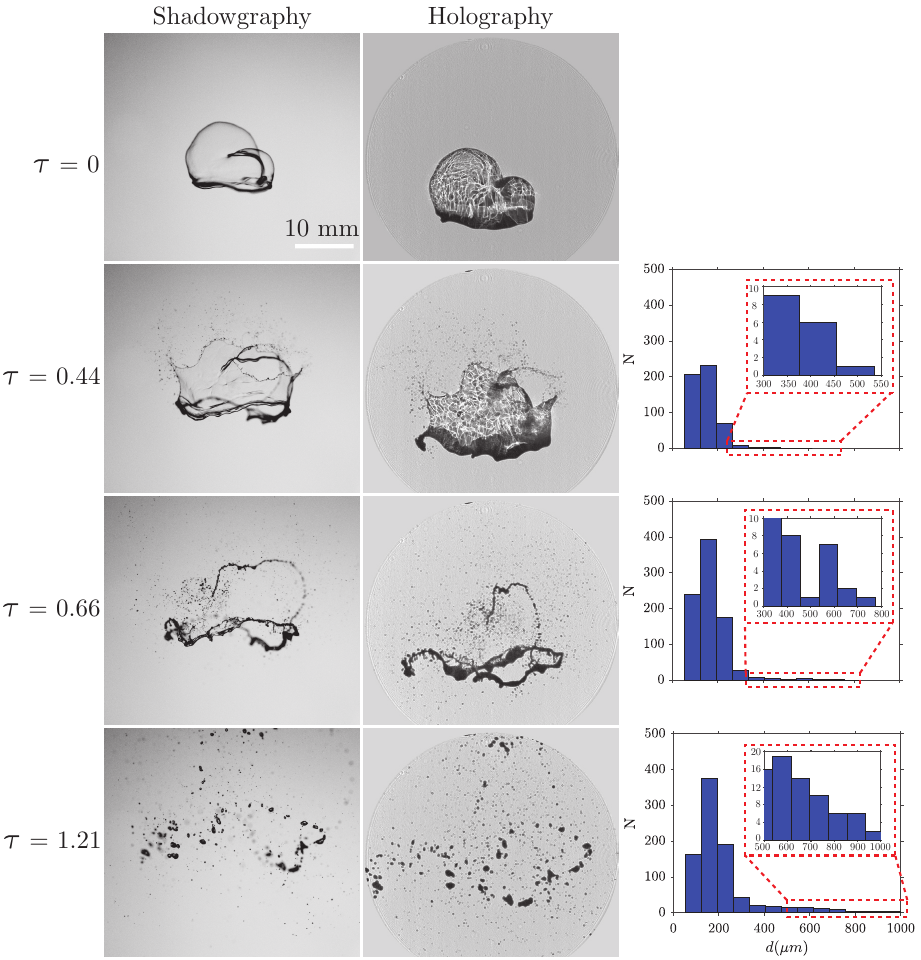}
\caption{Temporal evolution of the droplet size distribution for $\We = 18.9$ (dual-bag breakup). The panels in the first and second columns represent the shadowgraphy images and recorded holograms (obtained from the in-line holography technique). The values of the dimensionless time, $\tau$, are mentioned on the left side of the first column, measured from the instant at the onset of the breakup. The scale bar is shown in the top panel of the first column. The panels in the third column depict the histograms of the droplet size distribution (the droplet counts, $N$, versus the droplet diameter, $d$, in $\mu$m) at different instants for $\tau>0$.}
\label{dualbag}
\end{figure}

The temporal evolution of the DSD resulting from the dual-bag breakup phenomenon at $\We = 18.9$ is illustrated in figure \ref{dualbag}. Unlike the cross-flow configuration \citep{ade2023size,joshi2022droplet, boggavarapu2021secondary}, where the parent bag ruptures first, followed by the fragmentation of the core bag, the opposed-flow configuration in the present study shows both bags rupturing simultaneously at $\tau = 0.44$, as depicted in figure \ref{dualbag}. This simultaneous rupture generates child droplets with diameters ranging from $50~\mu$m to $550~\mu$m, forming a single peak in the histogram. At $\tau = 0.66$, the nodes from both bags detach, and the ligament connecting the two bags breaks apart as the bags retract toward the rim. This results in an increase in the number of child droplets within the size range of $50~\mu$m to $800~\mu$m, as shown in figure \ref{dualbag}. In the enlarged view of the histogram at this instant, a smaller peak corresponding to larger droplets ($500~\mu$m to $800~\mu$m) is also observed, primarily resulting from the fragmentation of larger nodes. Subsequently, the toroidal rim fragments due to capillary instability, while the nodes on the rim detach and break apart under the influence of Rayleigh-Taylor instability. This leads to a broader size distribution of child droplets ($50~\mu\text{m} < d < 1000~\mu$m), as seen in the histogram at $\tau = 1.21$, corresponding to the post-fragmentation stage. Some smaller satellite droplets from earlier time steps escape the frame, causing a noticeable reduction in the peak for smaller droplets ($50~\mu\text{m} < d < 250~\mu$m) compared to the peak at $\tau = 0.66$. In the enlarged view of the histogram at $\tau = 1.21$, another distinct peak for larger droplets ($d > 500~\mu$m) emerges, primarily due to the fragmentation of the rim and nodes as the aerodynamic forces weaken.

\begin{figure}
\centering
\includegraphics[width=0.6\textwidth]{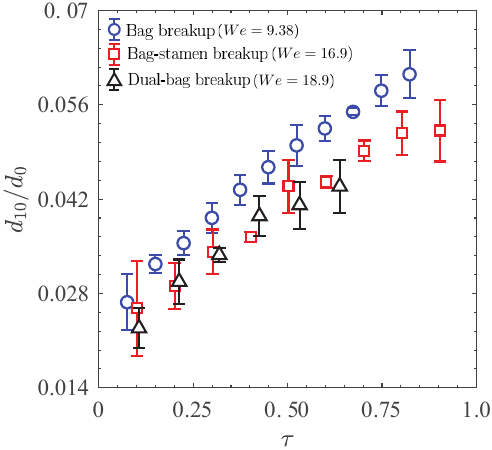}
\caption{Temporal variation of normalised number mean diameter ($d_{10}/d_{0}$) for different breakup processes. The error bar represents the standard deviation calculated from three repetitions of our experiments with the same set of parameters.}
\label{d10_figure}
\end{figure}

As discussed above, the size distribution of droplets based on droplet counts provides valuable information about the relative quantity of different droplet sizes. However, such distributions alone do not capture a representative measure that reflects the overall fragmentation dynamics. This limitation arises because size distributions emphasize the population of specific droplet sizes without accounting for the mean size trend or its evolution over time. The number mean diameter, $d_{10}$, is an important measure to bridge this gap, offering a quantitative measure that accounts for the average droplet size weighted by number density. The temporal variation of the normalized number mean diameter ($d_{10}/d_{0} = \int_{0}^{\infty } d p(d) \textrm{d}d / d_{0}$) for the bag, bag-stamen and dual-bag fragmentation processes is plotted in figure \ref{d10_figure}. Here, $p(d)$ is the probability density function of the diameter of the child droplets, $d$. It can be seen in figure \ref{d10_figure} that, at early times, $d_{10}/d_{0}$ is smaller for all three breakup cases. This is because the rupture of the thin bag initially creates only smaller fragments. At later times, the bag breakup case shows a larger value of $d_{10}/d_{0}$ due to the fragmentation of the rim and nodes, producing very large fragments. In contrast, for the bag-stamen breakup, $d_{10}/d_{0}$ is smaller at later stages compared to the bag breakup, as the breakup of the additional stamen reduces the contribution of larger-sized node droplets. It can also be observed that in the dual-bag breakup, $d_{10}/d_{0}$ is smaller than in the bag-stamen breakup, as the fragmentation of the two bags produces a greater number of smaller-sized fragments and fewer larger-sized rim and node droplets.

\subsection{Theoretical modeling} \label{sec:vpd}

In the previous section, we presented the droplet size distribution using count-based data ($N$ versus $d$). However, it is important to recognize that count-based data or number density overemphasizes smaller droplets due to their higher abundance. For example, during the bag breakup process, the rupture of the bag produces a large number of small droplets, whereas the rim and node fragmentation generate fewer larger droplets. While the count-based distribution effectively highlights smaller droplets, it provides limited insight into the contributions of rim and node droplets. Moreover, count-based distributions, although useful for quantitative insights, do not capture the full nature of the distribution, such as whether it is mono-modal, bi-modal, or multi-modal. To address these limitations and gain a more comprehensive understanding, analyzing the size distribution based on droplet volume is required. Thus, in this section, we examine the volume probability density ($P_v$), defined as the ratio of the total volume of droplets of a specific diameter to the total volume of all droplets. \citet{jackiw2022prediction} proposed an analytical model to predict the combined multi-modal size distribution for the aerodynamic breakup of droplets in a cross-flow configuration. In the present study, we adopt a similar approach to that of \citet{jackiw2022prediction} using volume probability density, which can be expressed as:

\begin{equation} \label{Pv}  
P_{v} = \frac{\zeta^3 P_{n}}{\int_{0}^{\infty} \zeta^3 P_{n} d\zeta} = \frac{\zeta^3 P_{n}}{\beta^3 \Gamma(\alpha+3)/\Gamma(\alpha)}.  
\end{equation}
Here, $P_n = \frac{\zeta^{\alpha -1} e^{-\zeta / \beta}}{\beta^{\alpha} \Gamma(\alpha)}$, \ks{wherein $\zeta=d/d_{0}$}; $\Gamma(\alpha)$ represents the gamma function; $\alpha = (\bar{\zeta}/\sigma_s)^2$ and $\beta = \sigma_s^2 / \bar{\zeta}$ are the shape and rate parameters, respectively; $\bar{\zeta}$ and $\sigma_s$ are the mean and standard deviation of the distribution. The droplet fragmentation process involves three main modes: bag rupture, rim fragmentation, and node breakup. It is crucial to account for the volume contribution from each mode. The shape and rate parameters of the gamma distribution can be calculated from characteristic sizes corresponding to each breakup mode \citep{jackiw2022prediction,ade2022droplet}. \ks{The fragmentation of bag, rim, and node breakup modes results in eleven characteristic sizes. Bag rupture specifically contributes four characteristic sizes associated with different instabilities. The mean $(\bar{\zeta})$ of the distribution is calculated as the arithmetic average of these four characteristic sizes, while the standard deviation $(\sigma_s)$ is determined as the square root of the variance of these sizes for bag breakup. Then, the shape parameter ($\alpha$) and rate parameter ($\beta$) of the gamma distribution for the bag rupture mode are calculated as $\alpha = (\bar{\zeta}/\sigma_s)^2$ and $\beta = \sigma_s^2 / \bar{\zeta}$. A similar methodology is applied to rim fragmentation and node breakup, which involve four and three characteristic sizes, respectively. The characteristic sizes corresponding to each breakup mode are used to compute $\bar{\zeta}$ and $\sigma_s$, which in turn define the corresponding $\alpha$ and $\beta$ values.}

Next, the overall size distribution of the breakup process can be predicted by determining the volume contributions of each mode. \ks{This involves first determining the volume contribution of each breakup mode, such as bag, rim, and nodes, relative to the total fragmented volume. The volume contribution of each mode is then multiplied by its respective volume probability density function, which characterizes the size distribution of that mode. Finally, summing these weighted distributions of all modes provides the overall drop size distribution (DSD).} In the subsequent section, we first describe the calculation of volume contributions, followed by the evaluation of the characteristic sizes resulting from the fragmentations of the droplet at different Weber numbers.

At moderate Weber numbers, the droplet deforms into a disk under the influence of the aerodynamic field, eventually evolving into a bag structure attached to a rim. The normalised deformed volume of the droplet ($V_D$) when exposed to an airstream can be expressed as \citep{jackiw2021aerodynamic}:
\begin{equation} \label{VD} 
\frac{V_{D}}{V_{0}}=\frac{3}{2}\left[ \left( \frac{2R_{i}}{d_{0}} \right)^{2}\left( \frac{h_{i}}{d_{0}} \right) - 2\left( 1 - \frac{\pi}{4} \right)\left( \frac{2R_{i}}{d_{0}} \right)\left( \frac{h_{i}}{d_{0}} \right)^{2} \right], 
\end{equation}
where $V_0$ is the initial droplet volume, $h_i$ represents the disk thickness, and $2R_i$ denotes the major diameter of the rim. The parameters $h_i$ and $2R_{i}$ can be evaluated using the following expressions \citep{jackiw2021aerodynamic, jackiw2022prediction}:
\begin{equation} 
\label{hi} \frac{h_{i}}{d_{0}}=\frac{4}{\We_{rim}+10.4}, \hspace{2mm} {\rm and} \hspace{2mm} \frac{2R_{i}}{d_{0}}=1.63-2.88e^{(-0.312\We)}, \end{equation}
where $\We_{rim} = {\rho_w \dot{R}^2 d_{0}/\sigma}$ denotes the rim Weber number, which characterizes the competition between the radial momentum at the droplet periphery and the restoring surface tension force. The constant radial expansion rate of the drop, $\dot{R}$, is given by $\dot{R}=(1.125 U \sqrt{\rho_a/\rho_w)}/2)(1-32/9 \We)$ \citep{jackiw2022prediction}. It is to be noted that for the bag breakup process, $V_{D}/V_{0} \approx 1$. This indicates that the entire droplet is deformed, leaving no undeformed core. A typical bag undergoes three distinct breakup modes: bag rupture, rim fragmentation, and node breakup. The volume contributions of the bag ($w_B$), rim ($w_{R}$), and node ($w_{N}$) breakup modes can be determined as follows \citep{jackiw2022prediction}:
\begin{eqnarray} 
w_B &=&\frac{V_{B}}{V_{0}} =\frac{V_{D}}{V_{0}}-\frac{V_{N}}{V_{0}}-\frac{V_{R}}{V_{0}}.  \label{wB} \\
w_{R} &=&\frac{V_{R}}{V_{0}}=\frac{3\pi }{2}\left [ \left ( \frac{2R_{i}}{d_{0}} \right )\left ( \frac{h_{i}}{d_{0}} \right )^{2}-\left ( \frac{h_{i}}{d_{0}} \right )^{3} \right ],\label{wR}  \\
w_{N} &=&\frac{V_{N}}{V_{0}}=\frac{V_{N}}{V_{D}}\frac{V_{D}}{V_{0}}, \label{wN}
\end{eqnarray}
Here, $V_N/V_D$ denotes the volume fraction of the node relative to the deformed droplet volume. \citet{jackiw2022prediction} experimentally studied bag and bag-stamen breakups across various Weber numbers and determined the mean value of $V_N/V_D$ to be 0.4. 

The next step is to determine the characteristic breakup size for each mode: bag, rim, and node. Using these characteristic sizes, the shape parameter ($\alpha$) and rate parameter ($\beta$) of the gamma distribution corresponding to each breakup mode can be calculated. In the following sections, we first present the characteristic breakup size for bag rupture, followed by those for rim fragmentation and node breakup.

\subsubsection{Bag rupture} 

Droplet fragmentation begins with the rupture of the bag, where the bag thickness ($d_B$) serves as the first characteristic size, expressed as \citep{culick1960comments}:
\begin{equation} \label{d_B} 
d_{B} = \frac{2\sigma}{\rho_w u_{rr}^2}, 
\end{equation}
where $u_{rr}$  represents the retraction velocity of the edge of the ruptured bag (receding rim). Upon rupture, the edge of the bag forms a receding rim that moves along the surface of the bag. In this study, $u_{rr}$ is determined experimentally by measuring the displacement rate of the receding rim relative to the main rim. 

The second characteristic size is related to the thickness of the receding rim ($d_{rr,B}$), which is defined as:
\begin{equation} \label{d_rr,B} 
d_{rr,B} = \sqrt{\frac{\sigma}{\rho_w a_c}}, 
\end{equation}
where $a_{c}=u_{rr}^2/R_{f}$ denotes the acceleration of the receding rim. Here, $R_f$ represents the radius of the bag at the time of its burst, and it can be evaluated as follow \citep{kirar2022experimental}:
\begin{equation} \label{Rf} 
R_f = \frac{d_{0}}{2\eta} \left [ 2e^{\tau^{\prime}\sqrt{p}} + \left ( \frac{\sqrt{p}}{\sqrt{q}}-1 \right )e^{-\tau^{\prime}\sqrt{q}} - \left ( \frac{\sqrt{p}}{\sqrt{q}}+1 \right )e^{\tau^{\prime}\sqrt{q}} \right ],
\end{equation}
where $\eta = f^2 - {120/\We}$, $p = f^2 - {96/\We}$, and $q = {24/\We}$. The stretching factor, $f$, is taken as $2\sqrt{2}$ \citep{kulkarni2014bag}. The dimensionless time, $\tau^{\prime}$, is given by $\tau^{\prime} = {Ut_b\sqrt{\rho_a / \rho_w}/d_0}$, where the bursting time, $t_b$, can be evaluated as \citep{jackiw2022prediction}:
\begin{eqnarray} \label{tb} 
t_b = \frac{\left [ \left ( \frac{2R_i}{d_0} \right ) - 2\left ( \frac{h_i}{d_0} \right ) \right ]}{\frac{2\dot{R}}{d_0}}  \times \hspace{2cm} \nonumber \\ 
\left [ -1 + \sqrt{1 + 9.4 \frac{8t_d}{\sqrt{3\We}} \frac{\frac{2\dot{R}}{d_0}}{\left [ \left ( \frac{2R_i}{d_0} \right ) - 2\left ( \frac{h_i}{d_0} \right ) \right ]} \sqrt{\frac{V_B}{V_0}}} \right ],
\end{eqnarray}
where $t_d = {d_0/U\sqrt{\rho_w / \rho_a}}$ is the deformation time scale.

The third characteristic size, $d_{RP,B}$, arises due to the Rayleigh-Plateau instability of the receding rim and is given by \citep{jackiw2022prediction}:
\begin{equation} \label{dRP,B} d_{RP,B} = 1.89 d_{rr,B}.
\end{equation}

\ks{The formation of ligaments from the edge of the receding rim of the bag undergoes nonlinear instabilities, leading to the generation of child droplets. The fourth characteristic size associated with these child droplets, $d_{sat,B}$, as given by \citep{keshavarz2020rotary}:}
\begin{equation} \label{dsat,B} 
d_{sat,B} = \frac{d_{RP,B}}{\sqrt{2 + 3Oh_{rr}/\sqrt{2}}}, 
\end{equation}
where $Oh_{rr} = {\mu_w/\sqrt{\rho_w d_{rr,B}^3 \sigma}}$ is the Ohnesorge number based on the receding rim thickness. Here, $\mu_w$ denotes the viscosity of water. The characteristic sizes, calculated using eqs. (\ref{d_B}), (\ref{d_rr,B}), (\ref{dRP,B}), and (\ref{dsat,B}), are utilized to estimate the mean and standard deviation of the droplet size distribution. These statistical measures are subsequently used to determine the parameters $\alpha$ and $\beta$ that define the gamma distribution (eq. \ref{Pv}) for the bag rupture mode.

\subsubsection{Rim fragmentation} 

Rim fragmentation initiates after the bag bursts, predominantly driven by the Rayleigh-Plateau instability. The first characteristic breakup size resulting from this instability is expressed as \citep{jackiw2022prediction}:
\begin{equation} \label{dR} 
d_R = 1.89 h_f,
\end{equation}
where $h_f = \sqrt{R_i / R_f}$ is the final rim thickness. Rim instability and fragmentation are also significantly influenced by collisions between the receding and main rims. The second characteristic size, $d_{rr}$, resulting from this collision, is given by \citep{jackiw2022prediction}:
\begin{equation} \label{drr} 
d_{rr} = d_0 \left[ \frac{3}{2} \left( \frac{h_f}{d_0} \right)^2 \frac{\lambda_{rr}}{d_0} \right]^{1/3},
\end{equation}
where $\lambda_{rr} = 4.5 d_{rr,B}$ is the wavelength of the receding rim instability.

The final mechanism involves the formation of child droplets from liquid ligaments near the pinch-off point, which influences both the receding rim collision and the Rayleigh-Plateau breakup. The characteristic sizes of these child droplets are given by \citep{keshavarz2020rotary}:
\begin{eqnarray} \label{dsatR} 
d_{sat,R} = \frac{d_R}{\sqrt{2 + 3 Oh_R / \sqrt{2}}}, \
d_{sat,rr} = \frac{d_{rr}}{\sqrt{2 + 3 Oh_R / \sqrt{2}}} \label{dsatrr},
\end{eqnarray}
where $Oh_R = {\mu_w/\sqrt{\rho_w h_f^3 \sigma}}$ is the Ohnesorge number based on the final rim thickness. Additional details can be found in Refs. \citep{jackiw2021aerodynamic,jackiw2022prediction,ade2022droplet}. The characteristic sizes obtained from Eqs. (\ref{dR}) - (\ref{dsatrr}) are used to calculate the mean and standard deviation, which then determine the values of $\alpha$ and $\beta$ for the distribution in the rim fragmentation mode.

\subsubsection{Node breakup} 

The breakup of nodes on the periphery of the rim is driven by Rayleigh-Taylor and Rayleigh-Plateau instabilities \citep{zhao2010morphological,kirar2022experimental}. The size, $d_{N}$, of the child droplets resulting from node breakup, is given by \citep{jackiw2022prediction}:
\begin{equation} \label{dN} 
d_{N} = d_{0} \left[ \frac{3}{2} \left( \frac{h_{i}}{d_{0}} \right)^{2} \frac{\lambda_{RT}}{d_{0}} n \right]^{1/3}, 
\end{equation}
where $\lambda_{RT} = 2\pi \sqrt{{3\sigma/\rho_{w} a}}$ is the maximum susceptible wavelength of the Rayleigh-Taylor instability, and
\begin{equation} \label{a} 
a = \frac{3}{4} C_{D} \frac{U^{2}}{d_{0}} \frac{\rho_{a}}{\rho_{w}} \left({D_{max} \over d_{0}} \right)^{2}, 
\end{equation}
is the acceleration of the deforming droplet. As suggested by \citet{zhao2010morphological}, the drag coefficient ($C_{D}$) of a disk-shaped droplet is approximately 1.2, and the extent of droplet deformation is given by ${D_{max} / d_0} = {2 / (1 + \exp{(-0.0019 \We^{2.7})})}$. In Eq. (\ref{dN}), $n = V_{N}/V_{D}$ represents the volume fraction of the node relative to the disk. \citet{jackiw2022prediction} estimated $n$ to be 0.2, 0.4, and 1, corresponding to the minimum, mean, and maximum sizes, respectively. These three values of $n$ define the characteristic sizes of the node droplets.

\begin{figure}
\centering
\includegraphics[width=0.95\textwidth]{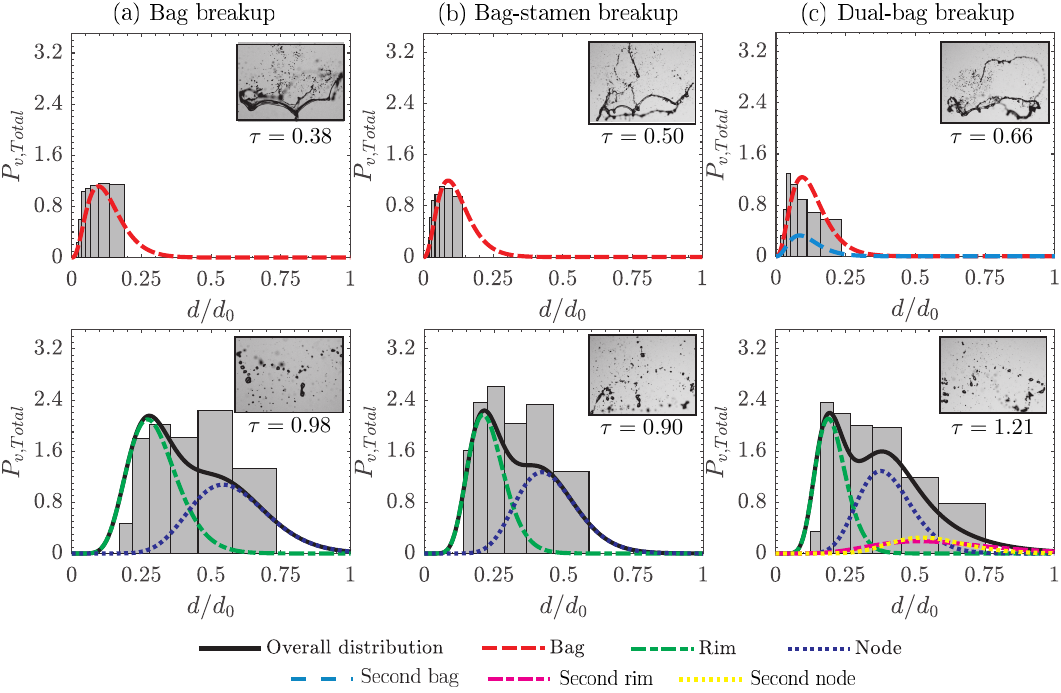}
\caption{Mode decomposition for (a) bag breakup at $\We = 9.38$, (b) bag-stamen at $\We = 16.9$, and (c) dual-bag breakup at $\We = 18.9$. The histograms represent the experimental results, and the theoretical predictions are depicted by lines. The inset in each panel presents the corresponding shadowgraph image of the fragmented droplet along with the dimensionless time instant.}
\label{fig1c}
\end{figure}

\subsubsection{DSD resulting from each breakup mechanism}
The contribution of each breakup mode (bag, rim, node, and undeformed core) to the overall size distribution (DSD) is evaluated theoretically using the parameters $\alpha$ and 
$\beta$, which defines the Gamma distribution, as discussed in the previous section. The individual distributions of each breakup mode are shown in figure \ref{fig1c}(a-c). The cases correspond to $\We = 9.38$, $\We = 16.9$, and $\We = 18.9$, representing normal bag, bag-stamen, and dual-bag fragmentation, respectively. \ks{The mode decomposition for bag breakup at $We = 9.38$, bag-stamen breakup at $We = 16.9$, and dual-bag breakup at $We = 18.9$, with standard deviation bars for each bin, obtained by averaging three experimental repetitions, is depicted in Figure S1 of the supplementary information.} Each panel of figure \ref{fig1c} depicts the results at two typical time instants during the fragmentation process, representing the earlier and later stages of fragmentation. The gray bars represent the experimental data, while the lines indicate the theoretical predictions.

It can be seen in figure \ref{fig1c}(a) that at $\tau = 0.38$, the bag fully ruptures, producing a size distribution characterized by a single peak, resulting in a mono-modal distribution. This observation aligns well with the theoretical prediction. By subtracting the droplets generated solely by bag fragmentation from the total child droplets at $\tau = 0.98$, the contributions of the rim and nodes are isolated, as shown in the second row of figure \ref{fig1c}(a). This results in a bi-modal distribution, with the larger peak corresponding to the rim and the smaller peak to node fragmentation, which is in agreement with the theoretical model. Quantitatively, the deformed volume of the droplet ($V_D$) accounts for $93.25\%$ of its initial volume ($V_0$) calculated from equation \ref{VD}. Among these, bag rupture contributes $16.63\%$, while the combined contributions of the rim and nodes amount to $76.62\%$.

Figure \ref{fig1c}(b) presents the mode decomposition for the bag-stamen breakup mechanism ($\We = 16.9$). At $\tau = 0.50$, the bag fully fragments, resulting in a mono-modal size distribution with a single peak consistent with the theoretical predictions. By subtracting the droplets produced solely by bag fragmentation from the total child droplets at $\tau = 0.90$, the contributions of the rim and nodes are isolated, as depicted in the second row of figure \ref{fig1c}(b). Similar to classical bag breakup, the distribution exhibits a bi-modal nature, with the larger peak corresponding to rim fragmentation and the smaller peak to node fragmentation, again aligning with the analytical model. The deformed volume of the droplet ($V_D$) constitutes $85.37\%$ of its initial volume ($V_0$). In this case, we observed that bag fragmentation contributes $16.38\%$, while rim and node fragmentation account for $34.84\%$ and $34.15\%$, respectively. Notably, the remaining $14.63\%$ of the undeformed core stretches into a stamen, which eventually fragments as well. \ks{This is because the local $\We$ experienced by the undeformed core is approximately 8.6, which is lower than the critical Weber number ($\We_{cr} = 9.38$) required for bag breakup, preventing the undeformed core from inflating into another bag.}

\begin{figure}
\centering
\includegraphics[width=0.9\textwidth]{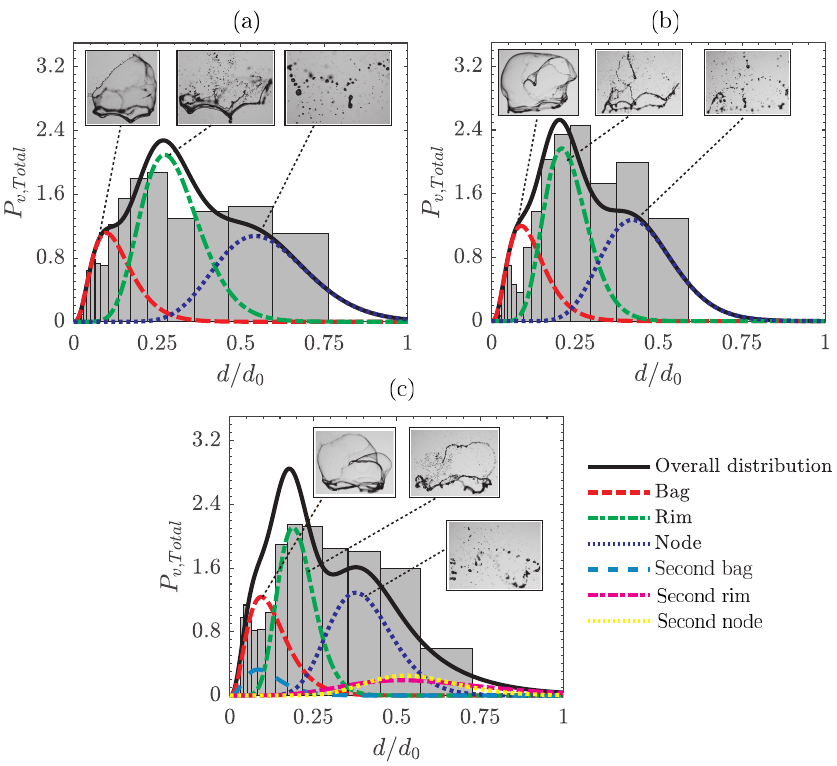}
\caption{Comparison of the overall size distribution obtained from the experiment with theoretical predictions for (a) $\We = 9.38$ at $\tau = 0.98$ (bag breakup), (b) $\We = 16.9$ at $\tau = 0.90$ (bag-stamen breakup), and (c) $\We = 18.9$ at $\tau = 1.21$ (dual-bag breakup). The solid line represents the overall size distribution, which is the sum of all modes. The individual contributions of each mode are also shown in each panel.}
\label{fig1d}
\end{figure}

In the dual-bag breakup mode at ($\We = 18.9$), about $77.56\%$ of the initial droplet volume ($V_0$) deforms, while the remaining portion persists as an undeformed core. The undeformed core, subjected to strong aerodynamic forces, subsequently inflates into a second bag and undergoes rupture. The second bag/core experiences a local Weber number, $\We = 13.24$, calculated using its equivalent diameter ($d_c$) derived from the core droplet volume, $V_c = V_0(1-V_D/V_0)$. This local Weber number exceeds the critical Weber number ($\We_{cr} = 9.38$), causing the core to deform, form a second bag, and rupture. Similar to the parent droplet, the core droplet undergoes three breakup modes (bag, rim, and node). The contribution of each mode, weighted by the volume of the core droplet, determines the child droplet distribution for the discrete modes. The combined size distribution of the second bag from the undeformed core ($P_{v,sb}$) is described by \citep{jackiw2022prediction}
\begin{equation} \label{b} 
P_{v,sb} = \frac{V_c}{V_0}(w_{N,sb}P_{v,sb,N} + w_{R,sb}P_{v,sb,R} + w_{B,sb}P_{v,sb,B}), 
\end{equation}
where the subscript (sb) stands for the second bag. $w_{N,sb} = V_{N,sb}/V_c$, $w_{R,sb} = V_{R,sb}/V_c$ and $w_{B,sb} = V_{B,sb}/V_c$ denote the contributions of volume weights from the second bag, rim of the second bag, and node of the second bag, respectively. $V_{N,sb}$, $V_{R,sb}$, and $V_{B,sb}$ are the volumes of the second bag, second rim, and nodes from the core droplet, respectively. The volume weights of each breakup mode for the second bag are determined in the same manner as for the first bag.

Figure \ref{fig1c}(c) shows the mode decomposition for the dual-bag breakup mechanism ($\We = 18.9$), with characteristic droplet sizes for each bag calculated using the same methodology as in the bag breakup case. At $\tau = 0.66$, both bags fragment simultaneously, unlike the cross-flow configuration where the parent bag bursts first, followed by the core bag. Since both bags rupture at the same time, their contributions are indistinguishable, resulting in a mono-modal size distribution with a single peak, as depicted in figure \ref{fig1c}. This observation aligns with the analytical model by \citet{jackiw2022prediction}. At $\tau = 1.21$, the contributions from the rim and nodes of both bags are isolated by subtracting the droplets produced by both bags from the total fragments. As shown in the second row of figure \ref{fig1c}(c), the first peak corresponds to fragments from the rim of the first bag. The second peak combines contributions from the nodes of the first bag and the bag, rim, and nodes of the second bag. Our experimental results, shown as histograms in figure \ref{fig1c}(c), clearly demonstrate this bi-modal size distribution. Quantitatively, the first bag, its rim, and its nodes account for $18.01\%$, $31.02\%$, and $28.52\%$ of the initial droplet volume ($V_0$), respectively. Meanwhile, the second bag, its rim, and nodes contribute $4.31\%$, $8.32\%$, and $9.81\%$, to the initial volume of the droplet, respectively.

\subsubsection{Overall size distribution}
In this section, we present the overall or combined size distribution from all breakup modes once the droplet breakup has ceased. The total volume probability density $(P_{v,Total})$ can be expressed as 
\begin{equation} \label{d} 
P_{v,Total} = P_{v,b} + P_{v,rim} + P_{v,nodes}, 
\end{equation} 
where $P_{v,b}$, $P_{v,rim}$, and $P_{v,nodes}$ represent the volume probability contributions from the bag, rim, and nodes, respectively.

Figure \ref{fig1d}(a) shows the overall distribution for the bag breakup case after fragmentation is complete at $\tau = 0.98$ (as shown in figure \ref{fig1a}). This distribution is characterized by three distinct peaks: the first at $d/d_0 \approx 0.12$ (bag), the second at $d/d_0 \approx 0.3$ (rim), and the third at $d/d_0 \approx 0.58$ (nodes). This indicates that the overall distribution is tri-modal in nature. This is consistent with observations by \citet{guildenbecher2017characterization, jackiw2022prediction, ade2023size} in cross-flow configurations.

Figure \ref{fig1d}(b) shows the overall volume probability distribution for the bag-stamen breakup at $\We = 16.9$ once fragmentation is complete at $\tau = 0.9$. Similar to typical bag breakup, in this case, the distribution displays three peaks: the first at $d/d_0 \approx 0.125$, the second at $d/d_0 \approx 0.23$, and the third at $d/d_0 \approx 0.34$. These peaks are due to the contributions from the bag, rim, node, and stamen from the undeformed core. The combined distribution displays a tri-modal nature, with the analytical model closely matching the experimental distribution trend. However, the contributions from the bag and node are overestimated and underestimated, respectively, due to the presence of the stamen. Additionally, an important observation is that the peaks corresponding to the bag and rim are closer together than in typical bag breakup fragmentation.

In the dual-bag breakup phenomenon ($\We = 18.9$), the total size distribution $(P_{v,Total})$ is the combination of the distributions from the first bag  $(P_{v,fb})$ and the second bag $(P_{v,sb})$. The distribution for the first bag is calculated using the same method as in the bag breakup case, while the second bag distribution is determined using equation \ref{b}. Thus, the combined distribution is expressed as,
\begin{equation} \label{f} 
P_{v,Total} = P_{v,fb} + P_{v,sb}. 
\end{equation} 
In the dual-bag breakup scenario, both the parent and core droplets undergo bag, rim, and node fragmentation, which collectively contribute to the overall size distribution. As depicted in figure \ref{fig1d}(c), the first peak at $d/d_0 \approx 0.1$ primarily results from the first and second bags, along with the rim of the first bag. The second peak at $d/d_0 \approx 0.4$ arises from the nodes of the first bag, as well as the rim and node associated with the second bag. Therefore, the overall size distribution displays a bi-modal nature, unlike the tri-modal distribution observed in single-bag fragmentation. Although the analytical model slightly overpredicts both peaks, it closely follows the experimental distribution trend. \ks{The overall droplet size distributions obtained from three experimental repetitions for $\We = 9.38$ (bag breakup mode), $\We = 16.9$ (bag-stamen breakup mode), and $\We = 18.9$ (dual-bag breakup mode), along with the corresponding standard deviations for each bin, are also depicted in Figure S2 of the supplementary information.}

\section{Concluding remarks}\label{sec:conc}

We investigate the morphology and breakup dynamics of freely falling droplets in a vertically moving airstream using shadowgraphy. A machine learning-based in-line holography technique is employed to analyze the size distribution of child droplets produced during various fragmentation processes for different Weber numbers. Additionally, Particle Image Velocimetry (PIV) experiments are conducted to examine the characteristics of the vertical airstreams. We observe that a droplet undergoes bag fragmentation at significantly lower Weber numbers in vertical airstreams (opposed-flow configuration, $\We \approx 6$) compared to that observed in a horizontal airstream (cross-flow configuration, $\We \approx 12$). The key difference between these configurations lies in the duration of droplet interaction with the potential core region of the airstream. In cross-flow configurations, a droplet travels through the core region rapidly, whereas, in vertical airstreams, the droplet remains within the core throughout the fragmentation process. This extended interaction with the counter-current airstream significantly alters the fragmentation dynamics, impacting the size distribution of the resulting child droplets for different Weber numbers. To the best of our knowledge, the characterization of the resultant child droplets produced due to the fragmentation of a primary droplet conducted in this study has not been explored yet despite its relevance to a wide range of applications, including industrial processes, rainfall estimation, combustion, surface coating, pharmaceutical production, disease transmission modelling, artificial rain technology, and many others.

Our study reveals that the interaction of a freely falling droplet with a vertically upward-moving airstream produces distinct fragmentation phenomena: bag, bag-stamen, and dual-bag breakup, observed at Weber numbers $\We = 9.38$, 16.9, and 18.9, respectively. At $\We = 9.38$, aerodynamic forces cause the droplet to undergo bag rupture in the early breakup stage, producing tiny child droplets. As the bag ruptures, a receding rim forms along the edge of the ruptured membrane, retracting and eventually merging with the primary rim. The Rayleigh-Plateau instability destabilizes the primary rim, creating smaller droplets, while the Rayleigh-Taylor instability drives the fragmentation of finger-like nodes, generating larger child droplets. At $\We = 16.9$, increased aerodynamic and shear forces lead to bag-stamen fragmentation. This mechanism resembles bag breakup but is characterized by the formation of an undeformed, stamen-like liquid bulb due to non-uniform deformation of the droplet. Both bag and bag-stamen breakups produce tri-modal size distributions; however, bag-stamen fragmentation generates fewer tiny droplets during the initial bag rupture. In later stages, the fragmentation of the rim, nodes, and stamen creates intermediate and larger droplets. At $\We = 18.9$, a dual-bag breakup occurs, exhibiting unique behaviour. Unlike cross-flow configurations, where dual-bag fragmentation typically happens at much higher Weber numbers ($\We \approx 35$) \citep{ade2023size,joshi2022droplet,boggavarapu2021secondary}, in opposed-flow at $\We = 18.9$, both bags inflate and burst simultaneously while the droplet remains within the airstream throughout the process. In this scenario, the resulting child droplets are characterized by a bi-modal size distribution, driven by the rupture of the two bags, rim and node fragmentation. A theoretical analysis using the two-parameter gamma distribution model proposed by \citet{jackiw2022prediction} effectively predicts the experimentally observed size distributions of child droplets for different Weber numbers. Thus, the present study enhances the understanding of vertical opposed-flow droplet breakup dynamics, paving the way for optimized numerical models for industrial and meteorological applications.

\vspace{2mm}

\vspace{2mm}
\noindent{\bf Acknowledgement:}
L.D.C. and K.C.S. express their gratitude to the Science \& Engineering Research Board, India, and Indian Institute of Technology Hyderabad, India for financial support through grants SRG/2021/001048 and IITH/CHE/F011/SOCH1, respectively. S.S.A. also acknowledges the support provided by the PMRF Fellowship.

%


\vspace{2mm}


\end{document}